\documentclass[10pt,draftclsnofoot,onecolumn,journal]{IEEEtran}
\usepackage[utf8]{inputenc}
\usepackage[pdftex]{graphicx}
\usepackage{textcomp}
\usepackage{pdfpages}
\usepackage{multirow}
\usepackage{listings}
\usepackage{float}
\usepackage[caption=false]{subfig}
\usepackage{graphics}
\usepackage{array}
\usepackage{color}
\usepackage{midfloat}
\usepackage[cmex10]{amsmath}
\usepackage{moreverb}
\usepackage{cite}
\usepackage{fixltx2e}
\usepackage{url}
\usepackage{fancyhdr}
\usepackage{rotating}
\usepackage{algorithmic}
\usepackage{algorithm}
\usepackage{paralist}
\usepackage{amssymb}
\usepackage{array}

\begin{document}

\title{Unveiling Potential Failure Propagation Scenarios in Core Transport Networks}

\author{\IEEEauthorblockN{Marc Manzano, Anna Manolova Fagertun, Sarah Ruepp, Eusebi Calle, \\Caterina Scoglio, Ali Sydney, Antonio de la Oliva, and Alfonso Mu\~{n}oz}%
\thanks{Marc Manzano and Eusebi Calle are with University of Girona, Spain. Anna Manolova Fagertun and Sarah Ruepp are with Technical University of Denmark, Denmark. Caterina Scoglio is with Kansas State University, USA. Ali Sydney is with Raytheon BBN Technologies, USA. Antonio de la Oliva and Alfonso Mu\~{n}oz are with University Carlos III of Madrid, Spain. Corresponding author: Marc Manzano (email: mmanzano@eia.udg.edu - marcmanzano@ksu.edu).}}%

\maketitle \thispagestyle{plain}

\begin{abstract}
The contemporary society has become more dependent on telecommunication networks. Novel services and technologies supported by such networks, such as cloud computing or e-Health, hold a vital role in modern day living. Large-scale failures are prone to occur, thus being a constant threat to business organizations and individuals. To the best of our knowledge, there are no publicly available reports regarding failure propagation in core transport networks. Furthermore, Software Defined Networking (SDN) is becoming more prevalent in our society and we can envision more SDN-controlled Backbone Transport Networks (BTNs) in the future. For this reason, we investigate the main motivations that could lead to epidemic-like failures in BTNs and SDNTNs. To do so, we enlist the expertise of several research groups with significant background in epidemics, network resiliency, and security. In addition, we consider the experiences of three network providers. Our results illustrate that Dynamic Transport Networks (DTNs) are prone to epidemic-like failures. Moreover, we propose different situations in which a failure can propagate in SDNTNs. We believe that the key findings will aid network engineers and the scientific community to predict this type of disastrous failure scenario and plan adequate survivability strategies.
\end{abstract}


\section{Introduction\label{sec:introduction}}
In the last decade, the Information and Communication Technology sector has substantially increased its dependency on communication networks for both business and pleasure. Additionally, this dependency has increased even more with the inception of the myriad of new emerging technologies and services such as smart-cities, cloud computing, e-Health and the Internet of the Things. 

Backbone Transport Networks (BTNs) constitute the foundations of the aforementioned network-dependent applications and services. Traditionally, BTNs have been classified in two types:
\begin{inparaenum}[(a)]
	\item Static Transport Networks (STNs); and
	\item Dynamic Transport Networks (DTNs).
\end{inparaenum} 
Typically, BTNs are divided in three differentiated layers: Data Plane (DP), Control Plane (CP), and Network Management Plane (NMP). STNs are centrally controlled by a network management system (NMS) such that the operation is manual and predetermined. On the contrary, DTNs are architectures under Automatically Switched Optical Network (ASON)/Generalized Multi-Protocol Label Switching (GMPLS) CP (or similar) where the management plane is a facilitator, but the actual network control is automated (to some extent) and services are configured and managed via distributed intelligence. 

The advent of Software-Defined Networking (SDN) technologies such as OpenFlow might contribute to changing BTNs as we know them today, realizing the concept of SDN-controlled Transport Networks \cite{McKeown2008Openflow,mcdysan2013SDTNS}. Currently, the Open Networking Foundation (ONF) has a dedicated working group defining guidelines for applying SDN standards to transport networks. Following the work in this Standards Developing Organization (SDO), major manufacturers of transport network equipment currently offer SDN-based products \cite{infinera2013ots}. In addition, large-scale SDNTNs such as the B4 Google network have been deployed \cite{googleB4}. As traditional DTNs, SDNTNs consist of a data and control plane such that the CP can be centralized or distributed \cite{Gringeri2013SDNTN,distributed2013sdtns}. The SDN-controller is the main component of the CP, and the applications that run at the controller provide the CP functionalities. Since extensive work regarding the development of standards for SDTNs is ongoing, we strongly believe that next-generation transport networks will be SDN-enabled. 

Although BTNs play a pivotal role to ascertain performance and integrity of the aforementioned novel services, their ubiquity is often taken for granted. Recently, network failures of great significance have occurred, re-enforcing the need to take the possibility of such large and potentially catastrophic failures into consideration in the underlying network design \cite{Habib2013630}. According to the European Network and Information Security Agency (ENISA), in 2011 at least 51 severe outages of communication networks were reported in Europe, where each affected about 400,000 users of fixed and mobile Internet \cite{CCC2011}.

\begin{figure}
\centering
\includegraphics[scale=0.7]{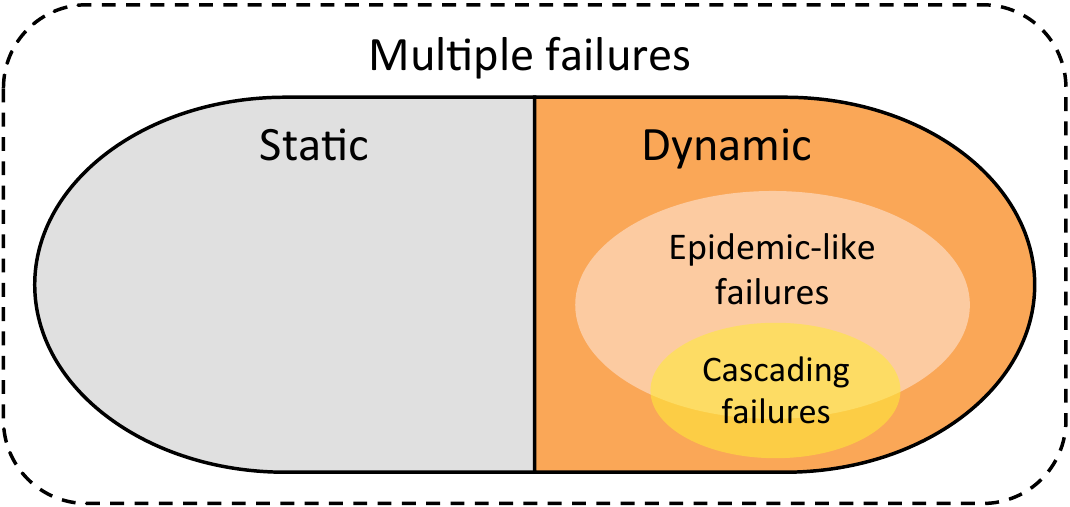}
\caption{Multiple failures broad classification. An epidemic-like failure is a process where a temporary failure propagates to physical neighbors. A cascading failure that triggers failures in physical neighbors is also considered as an epidemic-like failure.}
\label{fig:multfail}
\end{figure}

Many different protection and recovery techniques for single failures in communication networks have been extensively analyzed in the last decades. Consequently, in this work we focus on multiple failures, which can be broadly classified as depicted in Fig.~\ref{fig:multfail}. Multiple failures can be either \emph{static} or \emph{dynamic}. Static multiple failures are essentially one-off failures that affect nodes or links simultaneously at any given point (e.g., an earthquake). Dynamic failures have a temporal dimension. From all possible multiple failures, we focus on the dynamic scenarios and more specifically, in epidemic-like failure scenarios, where failure of one or more nodes might be propagated through the network, possibly resulting in an outbreak.

The aim of this work is to shed light on the following questions: \emph{``Are epidemic-like failures or attacks likely in BTNs?''} and \emph{``Will the upgrade to SDNTNs increase the vulnerability with respect to these type of failures or attacks?''} To do so, we present the state of the art of epidemic-like failure models in Section~\ref{soa}. Then, we review the main failure propagation model that has been proposed for transport networks in Section~\ref{epidemicsontelecom}. In Section~\ref{vulnerability} we discuss whether a failure propagation could occur in BTNs and SDNTNs. Section~\ref{providers} presents feedback from three network providers based on their experience in addition to the results of our research. Finally, Section~\ref{sec:conclusions} reviews the main contributions and findings of this work.

\section{What is an epidemic-like failure propagation?\label{soa}}

Epidemics theory has been used to describe and predict the propagation of diseases, human interactions, natural phenomena, and failures in a wide range of networks. An epidemic-like failure is a dynamic process where a partial/temporary failure propagates to physical neighbors. The spreading of the aforementioned events is formally represented by epidemic models, and can be generally classified in one of the following three families:
\begin{itemize}
	\item The \emph{Susceptible-Infected} (SI) considers individuals as being either susceptible (S) or infected (I). This family assumes that the infected individuals will remain infected forever, and so can be used for worst-case propagation scenarios ($S\rightarrow I$).
	\item The \textit{Susceptible-Infected-Susceptible} (SIS) considers that a susceptible individual can become infected on contact with another infected individual, then recovers with some probability of becoming susceptible again. Therefore, individuals might change their state from susceptible to infected, and vice versa, repeatedly ($S\leftrightarrows I$).
	\item The \textit{Susceptible-Infected-Removed} (SIR), which extends the SI model to take into account a removed state. Here, an individual can be infected just once because when the infected individual recovers, it becomes immune and will no longer pass the infection onto others ($S\rightarrow I \rightarrow R$).
\end{itemize}

As shown in Fig.~\ref{fig:multfail}, the subset of cascading failures intersects the subset of epidemic failures. Cascading failures are common in most critical infrastructures such as telecommunications, electrical power, rail, and fuel distribution networks \cite{Strogatz2001}. In telecommunication networks, we consider cascading failures as an epidemic when it occurs due to a malfunctioning in one node of a network which eventually triggers a failure in its neighbors. Real cascading failures in telecommunication networks have been observed in the IP layer of the Internet and in the physical layer of BTNs \cite{wrap32818}. 

The propagation of a cascading failure happens gradually in phases: after the initial failure (e.g., a massive broadcast of a routing message with a bug), some of the neighboring nodes get overloaded and fail. This first step leads to further overloading of more nodes and their collapse, constituting the second step and so on. In this way, networks go through multiple stages of cascading failures before they finally stabilize and there are no more failures. It is worth noting that cascading failures in other critical infrastructures, such as power grids, do not necessarily propagate by the physical contact of nodes or links, but by the load balancing in the global network. In such cases, cascading failures are not similar to epidemics, and thus are out of the scope of this work.

\section{A Failure Propagation Model for Telecommunication Networks\label{epidemicsontelecom}}

\begin{figure}
\centering
\includegraphics[scale=0.7]{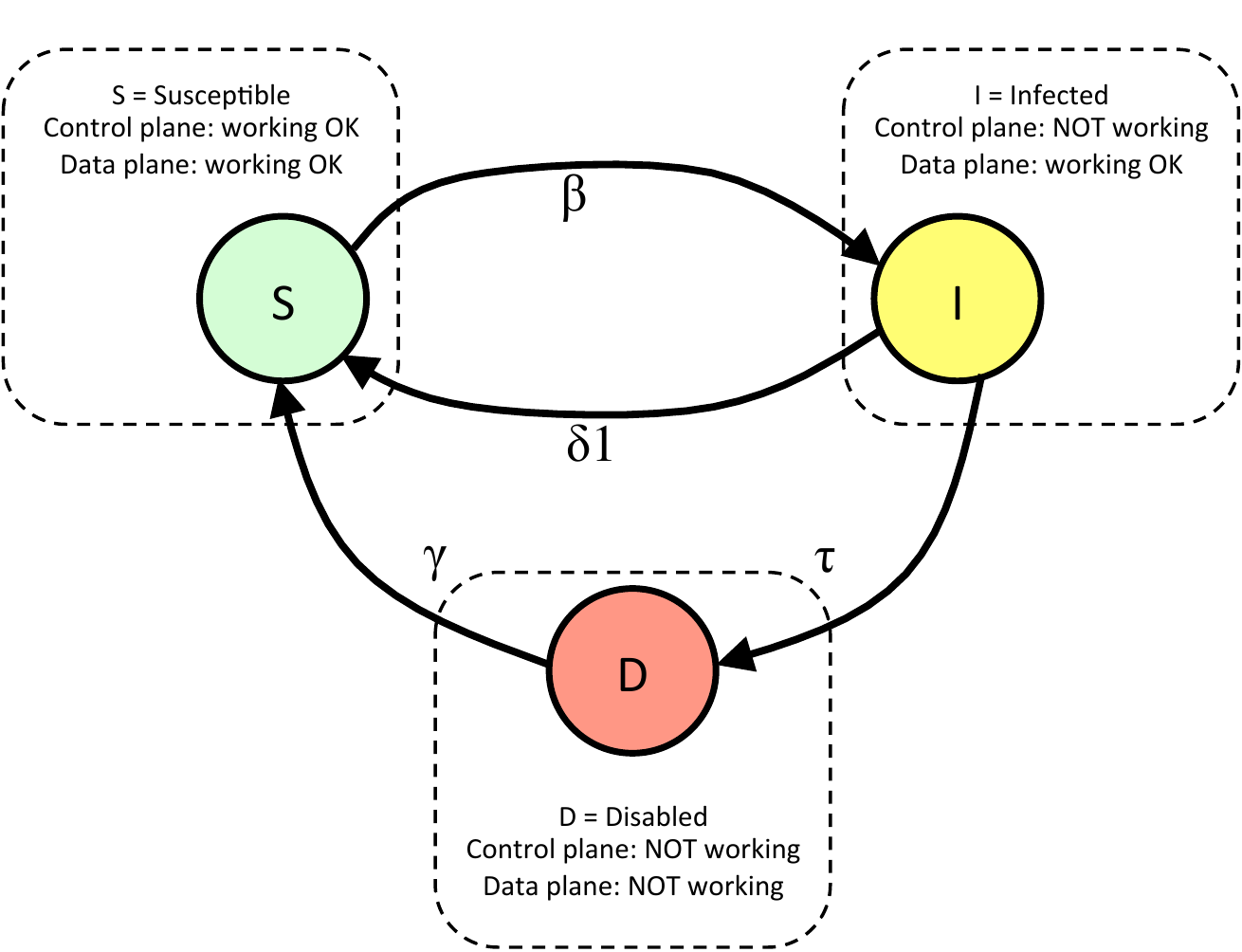}
\caption{State diagram of the SID model and its relationship with the STN and DTN planes.}
\label{fig:sid}
\end{figure}

Calle et al. pioneered the research in this field by presenting a new epidemic model called \emph{Susceptible-Infected-Disabled} (SID), which relates each state with a specific functionality of a node in the network \cite{calle2010multiple}. The state diagram of the SID model (\textit{Susceptible$\leftrightarrows$Infected$\rightarrow$Disabled$\rightarrow$Susceptible}), as seen from a single node, is shown in Fig.~\ref{fig:sid}. Each node, at each moment of time, can be either susceptible (S), infected (I) or disabled (D). A susceptible node can be infected with probability $\beta$ by receiving the infection from a neighbor (e.g., a bug in the routing or signaling protocol). An infected node can be repaired with probability $\delta_1$ (e.g., the network operator might manually reboot the CP). Finally, the disabled state takes into account the fact that the CP failure eventually affects the DP of the node with probability $\tau$ (e.g., the forwarding tables of the DP become unaccessible). After that, the model states that a repairing time, such as the mean time to repair (MTTR) of the node, determines when it becomes susceptible again ($\gamma$) (e.g., the required time to replace the node). 

No operations can be performed during control plane node failures
. However, as long as the data plane of the node does not fail, established connections should not fail or be re-routed as a result of control plane node failures. The routing protocol is assumed to be capable of detecting the failure of the control plane and informing all other nodes. Once the routing protocol has converged with this new information, a new connection will not be routed through this node. This same behavior is taken into account by the SID epidemic model.

\begin{figure}
 \centering
  \subfloat[]{
   \includegraphics[scale=0.55]{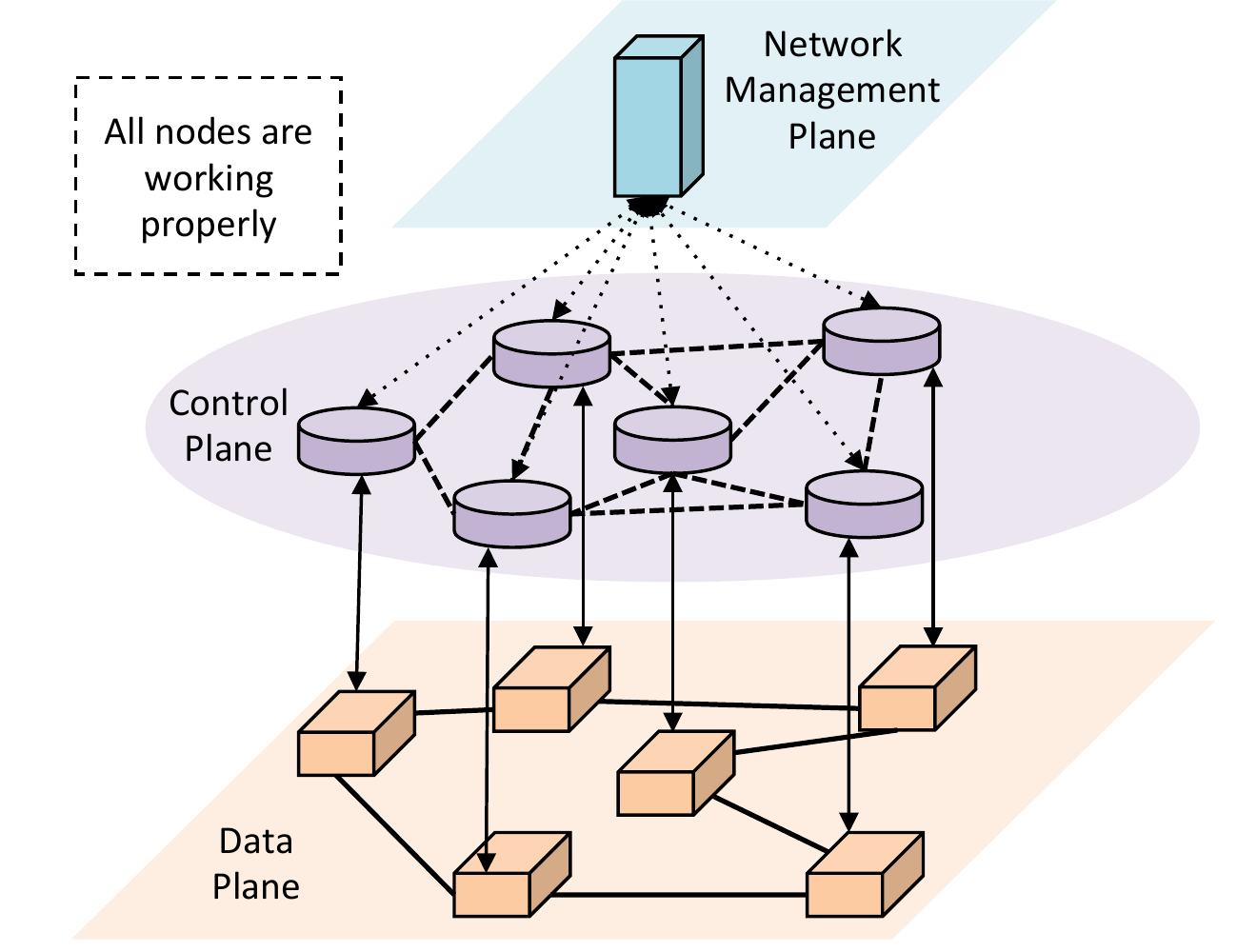}
    \label{fig:prop1}
  }
  \subfloat[]{
    \includegraphics[scale=0.55]{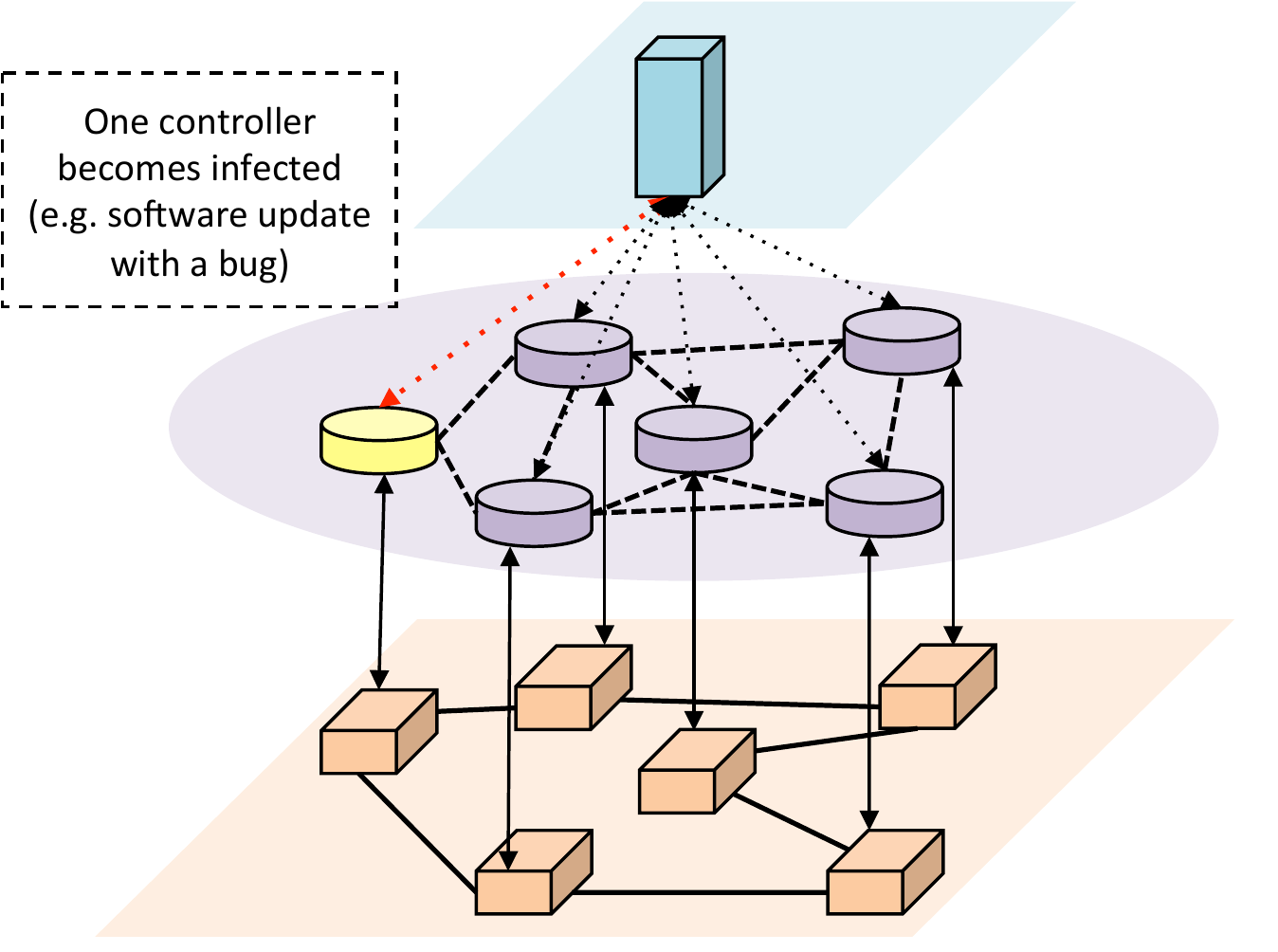}
    \label{fig:prop2}
  }
\\
\subfloat[]{
 \includegraphics[scale=0.55]{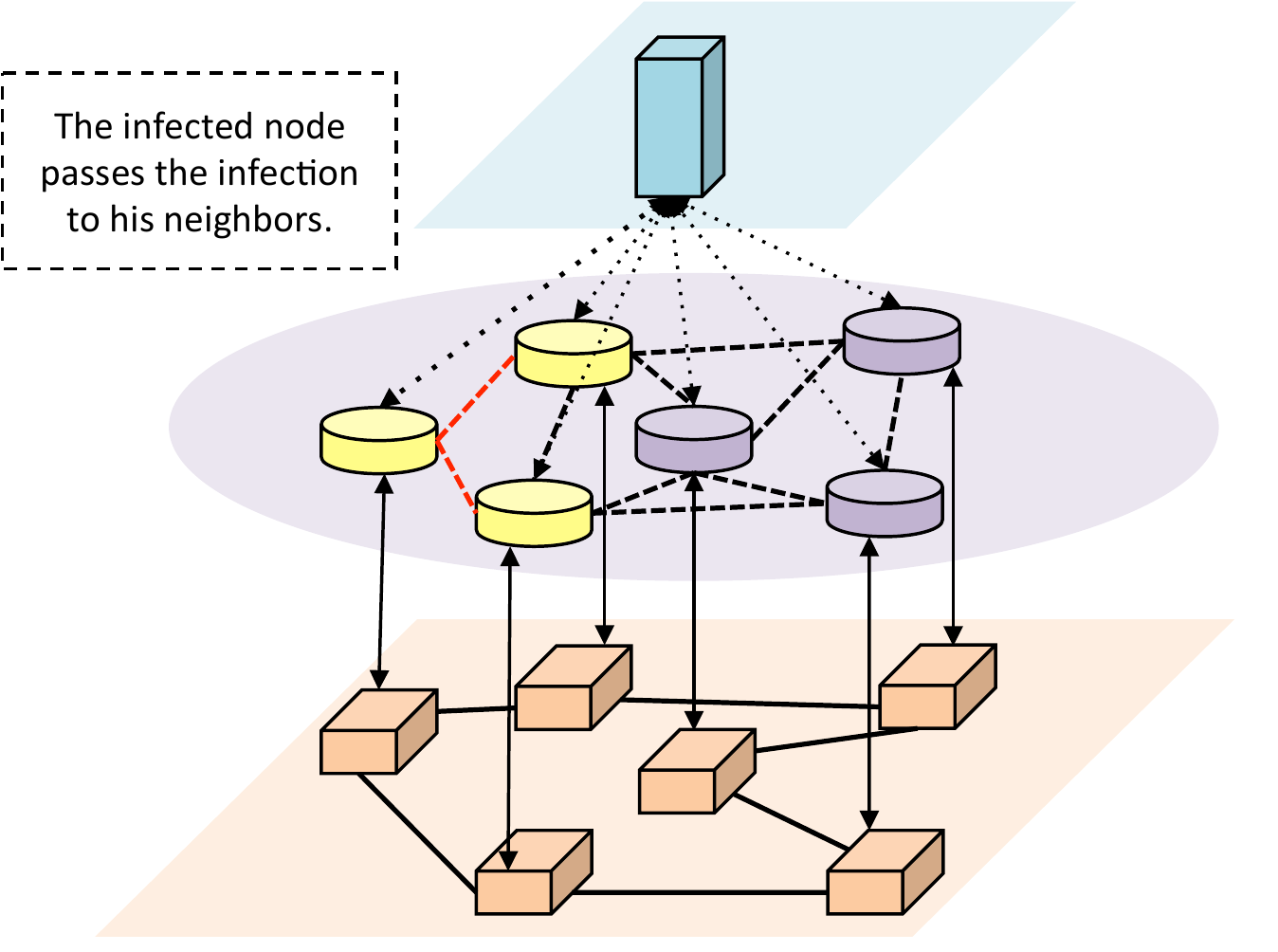}
  \label{fig:prop3}
}
\subfloat[]{
  \includegraphics[scale=0.55]{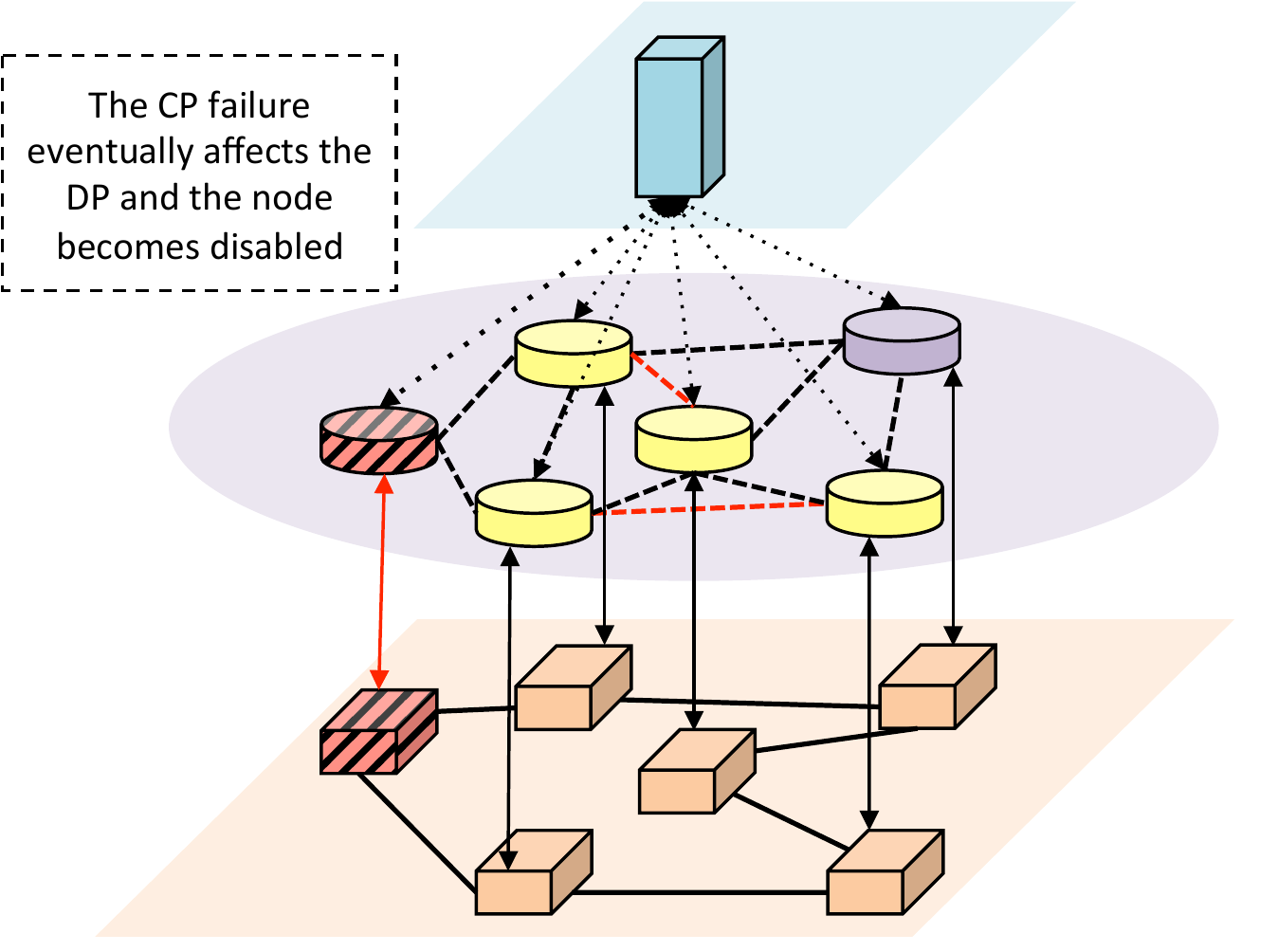}
  \label{fig:prop4}
}
  \caption{Example of how a failure can propagate in a BTN, according to the SID model.}
  \label{fig:failurepropagation}
\end{figure}

According to the SID model, Fig.~\ref{fig:failurepropagation} illustrates how a failure can propagate in a BTN. From Fig.~\ref{fig:prop1}, the network operates properly and thus, all nodes are in the susceptible state. At a given time $t$, the network management system updates a module of a controller with software that contains a bug, as shown in Fig.~\ref{fig:prop2}. As a result, this node becomes infected and propagates the failure (e.g., the bug) to his neighbors, as observed in Fig.~\ref{fig:prop3}. The epidemic continues to spread while the CP of an infected node eventually affects its DP operation. In this case, the incapability to resolve the problem at the CP might necessitate a complete node replacement (e.g., shutdown), thus impacting the operation of the DP. Consequently, this node becomes disabled as shown in Fig.~\ref{fig:prop4}.

Throughout the last decades, several failures have spread through communication networks. In the early 90s, a rapidly spreading malfunction collapsed the AT\&Ts long distance network\footnote{\url{http://users.csc.calpoly.edu/~jdalbey/SWE/Papers/att_collapse.html}}, causing the loss of more than \$60 million in terms of unconnected calls. In 2002, a failure propagation in the IP layer of the Internet was caused by a vulnerability of the BGP protocol. More recently, a BGP update bug which propagated through Juniper routers caused a major Internet outage in 2011\footnote{\url{http://www.zdnet.com/juniper-fail-seen-as-culprit-in-site-outages-4010024743/}}. In this latter case, routers were reseting and re-establishing its functioning state after five minutes.

Although there are no commercial references nor reports with respect to the occurrence of propagation of failures in BTNs, in the following sections we identify several failure scenarios that could be modelled as epidemic-like spreadings.


\section{Failure Propagation Scenarios in Transport Networks\label{vulnerability}}
This section describes BTNs and SDNTNs, showing that both contemporary and future networks might be predisposed to enduring epidemic-like failures.

\subsection{Backbone Transport Networks}

As explained in Section~\ref{sec:introduction}, a BTN is divided into Data Plane (DP), Control Plane (CP) and Network Management Plane (NMP), which have the following functionalities:
\begin{itemize}
	\item Data plane (or transport plane): responsible for user data transport, usually called data-path.
	\item Control plane: responsible for connection and resource management, which can be either associated with or separated from the managed DP.
	\item Management plane: responsible for supervision and management of the whole system (including transport and control planes).
\end{itemize}

As mentioned, there are two types of BTNs: static and dynamic. In STNs, the control plane does not act autonomously, but takes orders and updates from the NMP, which is eventually driven by a Network Management System (NMS). Thus, the intelligence of STNs is centralized. On the contrary, in DTN architectures such as GMPLS, the control plane dynamically reconfigures the network according to its current state. In this case, the NMP is only used as a facilitator (e.g. initializing or updating software of CP). In DTNs the control plane exists in each physical router and hence, it is distributed. For instance, GMPLS-based DTNs are networks where the control plane runs over an IP/Ethernet network, while the data plane runs over a wavelength routed WDM network.

Failures occurring in BTNs can be classified in two broad groups:
\begin{itemize}
	\item Faults: They include component failures as a result of natural exhaustion, human errors, and natural disasters.
	\item Attacks: They are intentional component failures. In fact, components may be selected to maximize the resulting impact of the attack. Furthermore, the identified targets may depend on various criteria such as the number of potentially affected users and additional socio-political and economic considerations.
\end{itemize}

There are three possible types of failures in the BTNs namely: (a) link; (b) node; and (c) software failures. In 2011 41\% of failures were attributed to hardware and software, 12\% to human errors, 12\% to natural disasters, and 6\% to malicious attacks \cite{CCC2011}. Although the percentage of malicious attacks was the smallest, they resulted in an average of 31 outage hours, as opposed to 17 outage hours for other failures. Therefore, despite the fact that targeted failures are not frequent, they are of paramount importance because they cause major disruptions. 

Control plane modules may fail due to software or hardware bugs, or protocol logic errors
. Recovery from such failures may involve switching to a hot standby, if a redundant software process has been implemented for that module. If the routing and signaling modules are implemented as separate processes, then either one of them may fail independently. If the signaling module fails, then new connections cannot be established through this node, and existing connections cannot be deleted.

In STNs, epidemic-like spreading of failures are rare because CP nodes do not interact between each other (i.e., they only communicate with the NMS). Therefore, if the NMS were compromised, a massive network failure affecting first the CP nodes and then the DP ones could easily occur (e.g., all nodes could be shut down). Nonetheless, this would not be an epidemic-like failure scenario, but a catastrophic static one. On the contrary, in DTNs an error during a software update, or a failure produced by a software itself could induce a CP node failure, which could spread through the network depending on the type of failure (e.g., a message with wrong objects in the fields). For instance, this initial failure, just like an infection, could be caused by the NMS. The spreading itself would be carried out between all nodes of the CP. This scenario is illustrated in Fig.~\ref{fig:failurepropagation}.

\subsection{SDN-controlled Transport Networks}


As with DTNs, SDNTNs require a DP and a CP. The SDN-controllers constitute the CP of a SDNTN and the network nodes constitute the DP. Applications that run at the controller provide the CP functionalities. Therefore, the SDN-controller can host different applications that support various carrier-class transport technologies such as SDH, DWDM, Ethernet or the suite of protocols of GMPLS.

\begin{figure}
\centering
\includegraphics[scale=0.65]{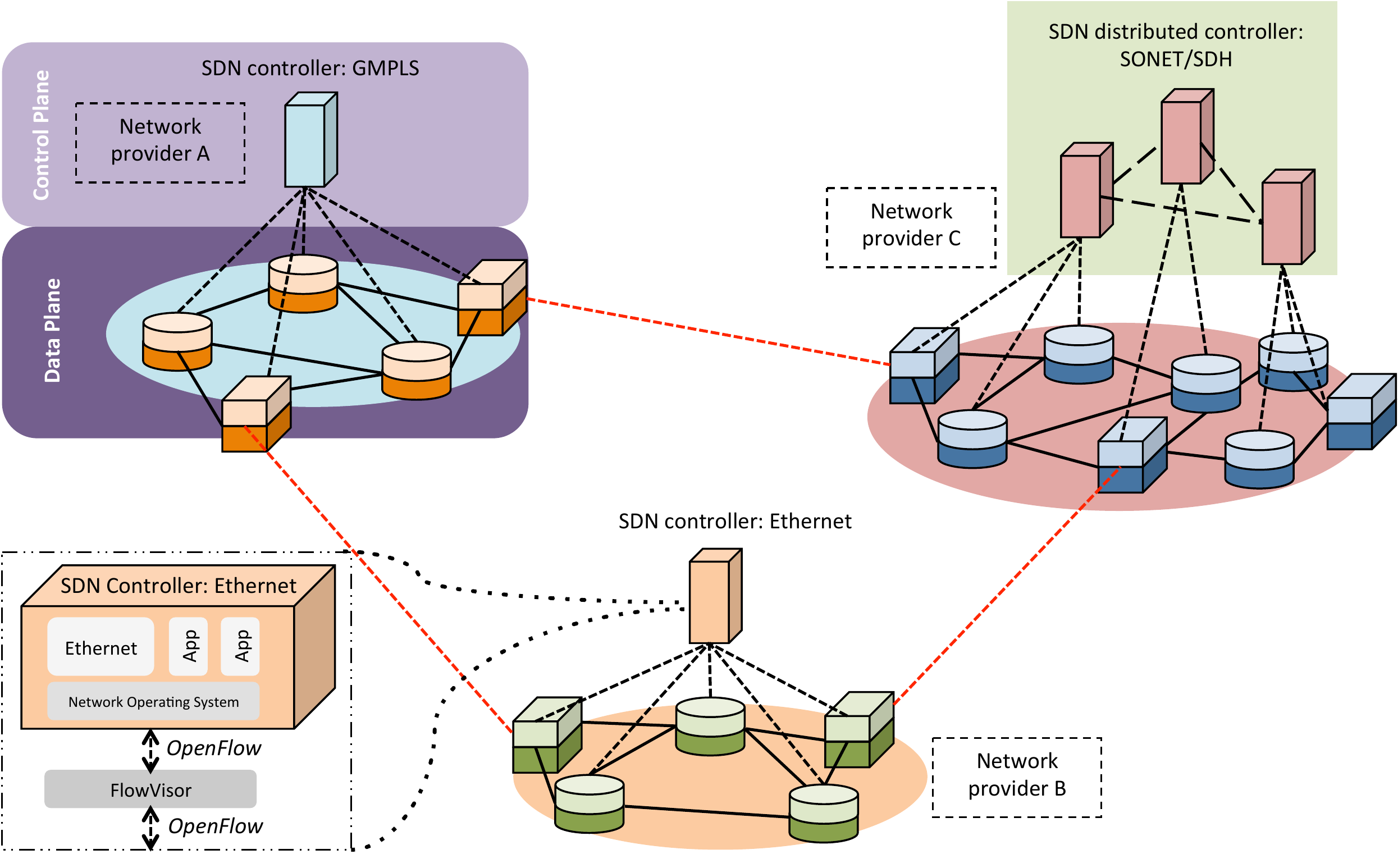}
\caption{Possible scenario of SDN transport networks.}
\label{fig:sdntn}
\end{figure}

Fig.~\ref{fig:sdntn} illustrates a possible scenario with three network providers operating SDNTNs. As observed, the CP can be either centralized (providers A and B) or distributed (provider C) \cite{Gringeri2013SDNTN,distributed2013sdtns}. As shown in the bottom left, the SDN-controller might host different applications that run on a Network Operating System (NOS). In addition, there is a component called FlowVisor that acts as a transparent proxy between the DP nodes and the SDN-controller. FlowVisor is a network virtualization layer and its main objective is to make sure that each user of the SDN-controller controls his own virtual network \cite{sherwood2009flowvisor}.

The emergence of SDNTNs implies an increased reliance on software. On the one hand, this aspect brings many benefits such as network programmability and control logic centralization. On the other hand however, it is also the source of major security concerns. It has been argued that SDN introduces new threat vectors with respect to traditional networks \cite{vectorsvulnerable2013SDN}. For instance, software faults hold a pivotal role in the reliability of SDNTNs, and significant efforts are made to define tools able to detect SW bugs or errors \cite{Canini2012NICE}. Additionally, other type of failures could occur due to policy (also called rules) conflicts, given that several users might work on the same physical network, each one on his own virtualized network. In such cases, FlowVisor is in charge to ensure compatibility among all policies.


\begin{figure}
 \centering

\subfloat[Vertical propagation]{
 \includegraphics[scale=0.55]{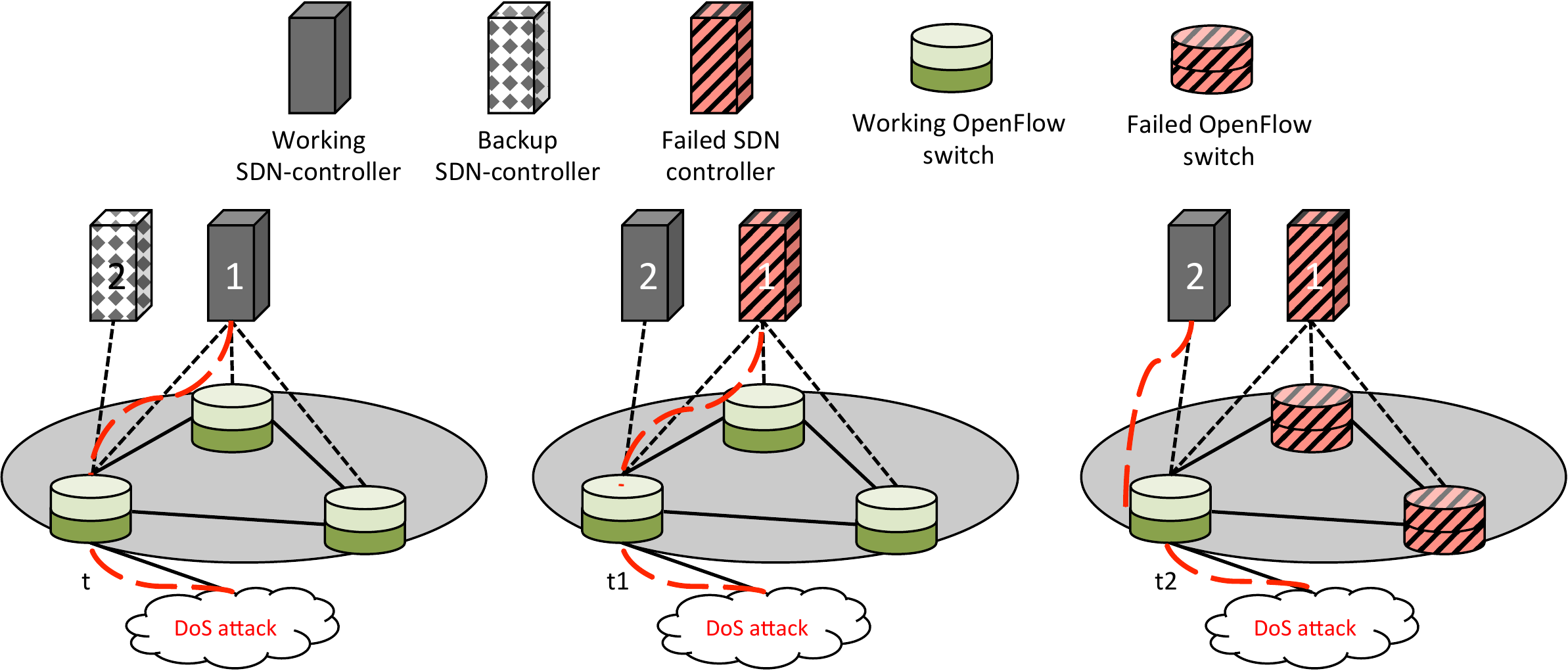}
  \label{fig:sdntn_prop2}
}
\\
  \subfloat[Horizontal DP propagation]{
   \includegraphics[scale=0.5]{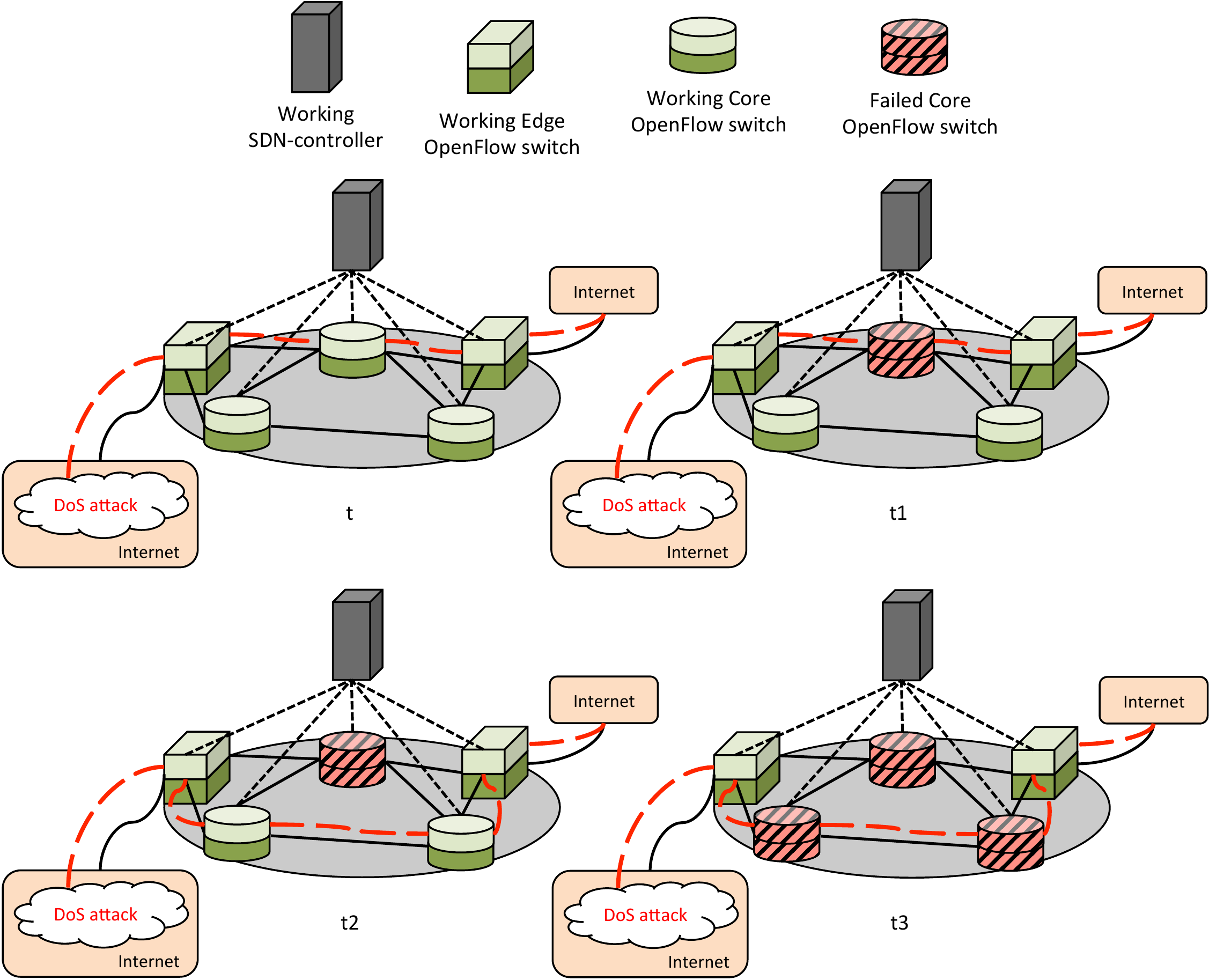}
    \label{fig:sdntn_prop1}
  }

  \caption{Example of failure propagation scenarios in SDNTNs. Successive increasing times are represented by $t$, $t1$, etc., to illustrate the propagation over time.}
  \label{fig:sdntn_prop}
\end{figure}

There are several cases in which epidemic-like failures could occur in SDNTNs. We classify these scenarios according to the initial event that triggers the propagation, which can be either a fault or an attack. First, in order to illustrate some potential scenarios in SDNTNs, we assume that a DDoS attack can be launched, for instance, by following the method provided in \cite{attacking2013SDN}. Consequently, we propose two propagation scenarios:
\begin{itemize}
	\item \emph{Vertical propagation}, which can be bottom-top or top-bottom: as shown in Fig.~\ref{fig:sdntn_prop2}, an attacker launches a DoS attack against an OpenFlow switch, which is connected to a primary SDN-controller. In order to increase the resilience of the system, the switch has been assigned a secondary SDN-controller. The switch needs to contact the SDN-controller for every packet. The SDN-controller is not able to handle all the new flow requests and fails. Then, the switch redirects its traffic to the secondary SDN-controller, where the same situation can happen. If the failing SDN-controller is the only controller available for other OpenFlow switches, then the switches lose their connection with the controller and can be considered as failed.
	\item \emph{Horizontal DP propagation}: as observed in Fig.~\ref{fig:sdntn_prop1}, an attacker injects huge quantities of traffic to an edge OpenFlow switch (e.g., a DoS attack). The destination of the injected traffic is outside the SDNTN, and to reach it, the traffic has to be routed through core switches and exit via another edge switch. Since SDN allows heterogeneous network scenarios (e.g., hardware from different vendors), we assume that edge OpenFlow switches have high-performance capabilities, while core OpenFlow switches are commodity devices. Therefore, the huge amount of injected traffic can eventually overflow the buffer of core switches, causing successive failures in the DP plane.
\end{itemize}
Second, it is possible to observe the same scenario shown in Fig.~\ref{fig:sdntn_prop1} caused by a fault. In such case, instead of being caused by a DoS attack, the switch failures would occur due to, for instance, a software bug in the routing protocol of the SDN-controller. Lastly, it is worth noting that the proposed horizontal DP propagation, as well as the vertical propagation, are closer to cascading failures. However, in the case of a distributed CP with several SDN-controllers, an epidemic-like spreading could happen if, for instance, the SDN-controllers where infected by masterful worms such as Tuxera\footnote{\url{http://spectrum.ieee.org/telecom/security/the-real-story-of-stuxnet}}.




\section{Transport Network Providers Experience with Large-Scale Failures\label{providers}}

With the purpose of validating some of the failure scenarios mentioned in the previous section, we have approached three network operators asking them about their experience with epidemic and cascading failure occurrence. The three operators were willing to discuss network vulnerabilities and specific spreading patterns within their networks. However, they were only willing to provide this information under a non-disclosure agreement. Consequently, we refer to the three operators as FO, SO and TO, which stand for first, second and third operator, respectively.

 

According to the experience of the FO with their STN, the most typical errors of hardware (HW) components are closely related to software (SW) bugs. Nonetheless, these failures have no connection to the outside world, i.e., no spreading is possible. The FO stated that their current procedures to operate their STN, which is based on the NMP, could potentially be vulnerable if the management server got compromised. This is highly unlikely, since the NMS has no access to the global Internet and a strong Firewall protection is provided. However, assuming a system breach, the damage to the network could be devastating, i.e., the entire network could fail simultaneously. As to epidemic-like failures, the only situation that the FO could see as problematic would be the SW update process. The SW update happens step-wise, where portions of the network are updated. Furthermore, when an element is updated, it has 2 banks of SW (i.e., an active one and a hot-standby). Typically, the server updates the hot-standby and then switches the operation of the element to that bank of SW, and then updates the primary one. If SW with a bug was loaded to the node, the node could stop operating. Since in STNs the node only communicates with the management server, no direct ``infection'' to the neighbors can occur. Nevertheless, considering that an entire region is updated (possibly with the same erroneous SW), then several elements can potentially be hit.

The SO uses equipment from a different vendor for their STN, and his experience with that equipment is different than the FO. In general, node HW failures are negligible in comparison with fiber cuts. The SO also suggested that SW updates might cause unresponsive management entity on the node. Furthermore, according to the SO, there is another potential vulnerability: the Digital Communication Network (DCN) interconnecting all nodes in a STN for the purpose of management and configuration information exchange. Typically, this case is related to standard packet-switched network problems, which are well-known from the IP world. Nonetheless, the SO outlined one possible problem: if buggy SW/mis-configurations were installed/occurred in the main routers (the designated router for example), as a result, nodes could start flapping the routes they knew from one port to the other. If one node became confused, he could advertise wrong routes to the rest and if flapping occurred, potentially blocking of the controller entities would be possible. This same scenario could be spread to the rest of nodes, since they also would need to readjust their router interfaces.

The TO has an extensive experience with GMPLS control plane technologies. This expert asserts that in STNs an epidemic spreading of failures is possible only if SW elements are failing, and if some type of distributed network (such as DCN) is deployed between nodes. On the contrary, in DTNs, where GMPLS is assumed to be one of the control plane solutions for an automatic provisioning and operation, this picture might change. The TO stated that there are numerous examples of spreading of failures in the contemporary IP distributed networks (e.g., based on BGP update bugs). Moreover, since the GMPLS CP is mostly a distributed IP-based network, it leads to the expectation that epidemic-like failures become reasonably possible. Three examples were outlined:
\begin{enumerate}
	\item When a node is affected by a CP failure and it attempts to recover its RSVP-TE state, synchronization becomes very difficult. This process involves synchronization with the neighbors, with the DP and with the NMP. Any of these three steps might fail for different reasons and this can directly impact the knowledge and states of the neighboring nodes. Such a process pertains mainly to the CP and possibly would affect only some of the connections. However, the fact is that an unsuccessful process of synchronization might lead to nodes operating in a state which is erroneous and potentially influencing the node's service delivery capability. 
	\item If a node fails, the neighbors of a node must update their state and the state of the network. In time, the neighbors will have moved towards a state they are not ``used to be'', i.e., an untypical state (the typical state is when everything is properly functioning). Under this situation, the neighbors might move closer to an unstable state of operation (e.g., if a node switches from a functioning to non-functioning state and vice versa) and this can potentially bring these nodes into a dysfunctional state. 
	\item Carrier-class components are typically undergoing very stringent tests and the requirements for robustness are extensive. Nevertheless, it is still possible to observe implementations where a failure in one single process might bring other processes in the same node to stop functioning. For example, this can occur if processes (routing and signaling) share CPUs (which might suffer HW/SW problems) or if they share memory (buffer overflows). 
\end{enumerate} 
Finally, the TO indicated that RSVP-TE is highly vulnerable to problems, and when problems occur, recovering the functional state in the network is much more difficult (due to the synchronization process described earlier). OSPF (Open Shortest Path First) failures are simpler to fix and synchronization between nodes might be established in a shorter period of time. Nevertheless, mis-configured routing is one of the main sources of large-scale failures in networks, as stated in the previous sections of this work. Thus, OSPF is also considered as a potential vulnerability.




\section{Conclusions\label{sec:conclusions}}
In this paper, we have studied the possibility of observing epidemic-like failures in Backbone Transport Networks and in SDN-controlled Transport Networks. We have analyzed both architectures and related each of them to several failure scenarios. Finally, to reinforce our study, three network operators have revealed several vulnerabilities in STNs and DTNs that could eventually lead to epidemic-like failures.

One of the most significant differences between traditional BTNs and SDNTNs is that BTNs are closed systems, so the surface of potential risks that might initiate the propagation of a failure is smaller that the one of SDNTNs, where homogeneous scenarios prevail. In any case, human-induced errors are the major issue, which could be either deliberate or unintentional. For this reason, the robustness of the CP of both DTNs and SDNTNs depends highly on the applied SW engineering principles, and on the way the processes and protocols are configured. Poor network management or mis-configured controllers, or inconsistent protocol implementations might become the root cause of epidemic-like failures in the CP. Additionally, cyber-security plays a pivotal role in DTNs and SDNTNs, given that in such networks the operation of each node relies exclusively on software. Since one of the potential threats is to compromise either the network management plane or the SDN-controller, it will be a matter of how to hack a closed system to gain non-granted access. To conclude, we anticipate that this paper will inspire researchers and professionals to design and implement security mechanisms to enhance the resilience of transport networks under dynamic multiple failure scenarios.

\section*{Acknowledgements}

This work is partially supported by Spanish Ministry of Science and Innovation project TEC 2012-32336, and by the Generalitat de Catalunya research support program SGR-1202. This work is also partially supported by the Secretariat for Universities and Research (SUR) and the Ministry of Economy and Knowledge through AGAUR FI-DGR 2012 and BE-DGR 2012 grants (M.~M.)

\bibliography{epidemicsOTN}
\bibliographystyle{IEEEtran}

\end{document}